\documentclass[11pt]{article}
\usepackage{amsfonts}
\usepackage{latexsym,amsmath}
\usepackage{amssymb,array}
\usepackage{calc}
\RequirePackage[dvips]{graphicx}
\usepackage{graphics}
\usepackage[colorlinks=true]{hyperref}
\hypersetup{urlcolor=blue, citecolor=red}
\makeatletter \oddsidemargin 0in \evensidemargin 0in \textwidth
16cm\textheight 20cm 
\begin{document}
\newcommand{\la}{\lambda}
\newcommand{\eq}{\Leftrightarrow}
\newcommand{\mf}{\mathbf}
\newcommand{\ri}{\Rightarrow}
\newtheorem{t1}{Theorem}[section]
\newtheorem{d1}{Definition}[section]
\newtheorem{n1}{Notation}[section]
\newtheorem{c1}{Corollary}[section]
\newtheorem{l1}{Lemma}[section]
\newtheorem{r1}{Remark}[section]
\newtheorem{e1}{Counterexample}[section]
\newtheorem{p1}{Proposition}[section]
\newtheorem{cn1}{Conclusion}[section]
\renewcommand{\theequation}{\thesection.\arabic{equation}}
\pagenumbering{arabic}
\title {Stochastic Comparison of Parallel Systems with Log-Lindley Distributed Components}
\author{Shovan Chowdhury\\Indian Institute of Management, Kozhikode\\Quantitative Methods and Operations Management Area
\\Kerala, India \and Amarjit Kundu\\Santipur College\\Department of
Mathematics\\ West
Bengal, India
}\maketitle
\begin{abstract}
In this paper, we study stochastic comparisons of parallel systems having log-Lindley distributed components. These comparisons are carried out with respect to reversed hazard rate and likelihood ratio ordering.
\end{abstract}
\section{Introduction}
\setcounter{equation}{0}
\hspace*{0.3in} In reliability optimization and life testing experiments, many times the tests are censored or truncated when failure of a device during the warranty period may not be counted or items may be replaced after a certain time under a replacement policy. Moreover, many reliability systems and biological organism including human life span are bounded above because of test conditions, cost or other constraints. These situations result in a data set which is modeled by distributions with finite range (i.e. with bounded support) viz. power function density, finite range density,  truncated Weibull, beta, Kumaraswamy and so on (see for example, Ghitany~\cite{gh}, Lai and Jones~\cite{lai1}, Lai and Mukherjee~\cite{lai2}, Moore and Lai~\cite{moo} and Mukherjee and Islam~\cite{muk}). 
\\\hspace*{0.3in} Recently, G$\acute{o}$mez et al.~\cite{go1} introduce the log-Lindley (LL) distribution with parameters $(\sigma,\lambda)$, written as LL($\sigma,\lambda$), as an alternative to the beta distribution with the probability density function given by
\begin{equation}\label{e0}
f(x;\sigma,\lambda)=\frac{\sigma^2}{1+\lambda\sigma}\left(\lambda-\log x\right) x^{\sigma-1};~0<x<1,~\lambda\geq 0,~\sigma>0,
\end{equation} 
where $\sigma$ is the shape parameter and $\lambda$ is the scale parameter. This distribution with a simple expression and nice reliability properties, is derived from the generalized Lindley distribution as proposed by Zakerzadeh and Dolati~\cite{za}, which is again a generalization of the Lindley distribution as proposed by Lindley~\cite{li}. The LL distribution exhibits bath-tub failure rates and has increasing generalized failure rate (IGFR). This distribution has useful applications in the context of inventory management, pricing and supply chain contracting problems (see, for example, Ziya et al.~\cite{zia}, Lariviere and Porteus~\cite{lar1} and Lariviere~\cite{lar2}), where a demand distribution is required to have the IGFR property. Moreover, it has application in the actuarial context where the cumulative distribution function (CDF) of the LL distribution is used to distort the premium principle (G$\acute{o}$mez et al.~\cite{go1}). The LL distribution is also shown to fit rates and proportions data better than the beta distribution (G$\acute{o}$mez et al.~\cite{go1}).        
\\\hspace*{0.3in} Order statistics play an important role in reliability optimization, life testing, operations research and many other areas. Parallel and series systems are the building blocks of many complex coherent systems in reliability theory. While the lifetime of a series system corresponds to the smallest order statistic $X_{1:n}$, the same of a parallel system is represented by the largest order statistic $X_{n:n}$. Although stochastic comparisons of order statistics from homogeneous populations have been studied in detail in the literature, not much work is available so far for the same from heterogeneous populations, due to its complicated nature of expressions. Such comparisons are studied with exponential, gamma, Weibull, generalized exponential or Fr$\acute{e}$chet distributed components  with unbounded support. One may refer to Dykstra \emph{et al.}~\cite{dkr11},  Misra and Misra~\cite{mm11.1}, Zhao and Balakrishnan~(\cite{zb11.2}), Torrado and Kochar~\cite{tr11}, Kundu and Chowdhury~\cite{kun2}, Kundu \emph{et al.}~\cite{kun1}, Gupta \emph{et al.}~\cite{gu} and the references there in. Moreover, not much attention has been paid so far to the stochastic comparison of two systems having finite range distributed components. The notion of majorization (Marshall et al. [5]) is also essential to the understanding of the stochastic inequalities for comparing order statistics. This concept is used in the context of optimal component allocation in parallel-series as well as in series-parallel systems, allocation of standby in series and parallel systems, and so on, see, for instance, El-Neweihi et al.~\cite{el}. It is also used in the context of minimal repair of two-component parallel system with exponentially distributed lifetime by Boland and El-Neweihi~\cite{bo}. \\
\hspace*{0.3 in} In this paper our main aim is to compare two parallel systems in terms of reversed hazard rate order and likelihood ratio order with majorized scale and shape parameters separately, when the components are from two heterogeneous LL distributions as well as from the multiple outlier LL random variables. The rest of the paper is organized as follows. In Section 2, we have given the required notations, definitions and some useful lemmas which have been used throughout the paper. Results related to reversed hazard rate ordering and likelihood ratio ordering between two order statistics $X_{n:n}$ and $Y_{n:n}$ are derived in Section 3. 
%%%%%%%%%%%%%%%%%%%%%%%%%%%%%%%%%%%%%%%%%%%%%%%%%%%%%%%%%%%%%%%%%%%%%%%%%%%%%%%%%%%%%%%%%%%%%%%%%%%%%%%%%%%%%%%%%%%%%%%%%%%%%%%%%%%%%%%%%%%%
\\\hspace*{0.3 in} Throughout the paper, the word increasing (resp. decreasing) and nondecreasing (resp. nonincreasing) are used interchangeably, and $\Re$ denotes the set of real numbers $\{x:-\infty<x<\infty\}$. We also write $a\stackrel{sign}{=}b$ to mean that $a$ and $b$ have the same sign. %Further, by $a\stackrel{\rm def}=b$ we mean that $b$ is defined as $a$. 
For any differentiable function $k(\cdot)$, we write $k'(t)$ to denote the first derivative of $k(t)$ with respect to $t$. 
\section{Notations, Definitions and Preliminaries}
\hspace*{0.3 in} For an absolutely continuous random variable $X$, we denote the probability density function, the distribution function and the reversed hazard rate function by $f_X(\cdot), F_X(\cdot),$ and $\tilde r_X(\cdot)$ respectively. The survival or reliability
 function of the random variable $X$ is written as $\bar F_X(\cdot)=1-F_X(\cdot)$.
\\\hspace*{0.3 in} In order to compare different order statistics, stochastic orders are used for fair and reasonable comparison.
In literature many different kinds of stochastic orders have been developed and studied.
The following well known definitions may be obtained in Shaked and Shanthikumar~\cite{shak1}.
\begin{d1}\label{de1}
Let $X$ and $Y$ be two absolutely continuous random variables with respective supports $(l_X,u_X)$ and $(l_Y,u_Y)$,
where $u_X$ and $u_Y$ may be positive infinity, and $l_X$ and $l_Y$ may be negative infinity.
Then, $X$ is said to be smaller than $Y$ in
\begin{enumerate}
\item[(i)] likelihood ratio (lr) order, denoted as $X\leq_{lr}Y$, if 
$$\frac{f_Y(t)}{f_X(t)}\;\text{is increasing in} \,t\in(l_X,u_X)\cup(l_Y,u_Y);$$
\item[(ii)] hazard rate (hr) order, denoted as $X\leq_{hr}Y$, if $$\frac{\bar F_Y(t)}{\bar F_X(t)}\;\text{is increasing in}\, t \in (-\infty,max(u_X,u_Y)),$$
 which can equivalently be written as $r_X(t)\geq r_Y(t)$ for all $t$;
  \item[(iii)] reversed hazard rate (rhr) order, denoted as $X\leq_{rhr}Y$, if $$ \frac{F_Y(t)}{ F_X(t)}\;\text{is increasing in}\, t \in(min(l_X,l_Y),\infty),$$
 which can equivalently be written as $\tilde r_X(t)\leq \tilde r_Y(t)$ for all $t$;
  \item[(iv)] usual stochastic (st) order, denoted as $X\leq_{st}Y$, if $\bar F_X(t)\leq \bar F_Y(t)$ for all \\$t\in (-\infty,\infty).$ 
\end{enumerate}
\end{d1}
In the following diagram we present a chain of implications of the stochastic orders, see, for instance, Shaked and Shanthikumar \cite{shak1}, where the definitions and usefulness of these orders can be found.
\vspace{0.17 in}
\\\hspace*{1.7 in}$~~~~~~X\leq_{hr}Y$
\\\hspace*{1.7 in}$~~~~~~~~~~~\uparrow ~~~~~~~\searrow$
\\\hspace{6 in} $~~~~~~~~~~~~~~~~~~~~~~~~~~~~~~~~~~~~~~~~~X\leq_{lr}Y~~\rightarrow~~X\leq_{st}Y.$

\hspace{2.51 cm}$~~~~~~~~~~~~~~~~~~~~~~~~~\downarrow~~~~~~~~~\nearrow$

\hspace{2 cm}$~~~~~~~~~~~~~~~~~~~~~~~~X\leq_{rhr}Y$
\\\hspace*{0.3 in} It is well known that the results on different stochastic orders can be established on using majorization order(s). Let $I^n$ denotes an $n$-dimensional Euclidean space where $I\subseteq\Re$. Further, let $\mathbf{x}=(x_1,x_2,\dots,x_n)\in I^n$ and $\mathbf{y}=(y_1,y_2,\dots,y_n)\in I^n$ be any two real vectors with $x_{(1)}\le x_{(2)}\le\cdots\le x_{(n)}$ being the increasing arrangements of the components of the vector $\mathbf{x}$. The following definitions may be found in Marshall \emph{et al.} \cite{Maol}.\\
\begin{d1}
%\begin{enumerate}
The vector $\mathbf{x} $ is said to majorize the vector $\mathbf{y} $ (written as $\mathbf{x}\stackrel{m}{\succeq}\mathbf{y}$) if
%\begin{equation*}\sum_{i=1}^j x_{[i]}\ge\sum_{i=1}^j y_{[i]},\;j=1,\;2,\;\ldots, n-1,\;\;and\; \;\sum_{i=1}^n x_{[i]}=\sum_{i=1}^n y_{[i]}.\end{equation*}
%or equivalently,
\begin{equation*}
\sum_{i=1}^j x_{(i)}\le\sum_{i=1}^j y_{(i)},\;j=1,\;2,\;\ldots, n-1,\;\;and \;\;\sum_{i=1}^n x_{(i)}=\sum_{i=1}^n y_{(i)}.
\end{equation*} 
%\item [(ii)] The vector $\mathbf{x}$ is said to weakly supermajorize the vector $\mathbf{y}$
 %(written as $\mathbf{x}\stackrel{\rm w}{\succeq} \mathbf{y}$) if
 %\begin{eqnarray*}
 % \sum\limits_{i=1}^j x_{(i)}\leq \sum\limits_{i=1}^j y_{(i)}\quad \text{for}\;j=1,2,\dots,n.
 %\end{eqnarray*}
 %\item [(iii)] The vector $\mathbf{x}$ is said to weakly submajorize the vector $\mathbf{y}$
 %(written as $\mathbf{x}\;{\succeq}_{\rm w} \;\mathbf{y}$) if
 %\begin{eqnarray*}
  %\sum\limits_{i=j}^n x_{(i)}\geq \sum\limits_{i=j}^n y_{(i)}\quad \text{for}\;j=1,2,\dots,n.
 %\end{eqnarray*}
%\end{enumerate}
\end{d1}

\begin{d1}
A function $\psi:I^n\rightarrow\Re$ is said to be Schur-convex (resp. Schur-concave) on $I^n$ if 
\begin{equation*}
\mathbf{x}\stackrel{m}{\succeq}\mathbf{y} \;\text{implies}\;\psi\left(\mathbf{x}\right)\ge (\text{resp. }\le)\;\psi\left(\mathbf{y}\right)\;for\;all\;\mathbf{x},\;\mathbf{y}\in I^n.
\end{equation*}
\end{d1}

\begin{n1}
Let us introduce the following notations.
\begin{enumerate}
\item[(i)] $\mathcal{D}_{+}=\left\{\left(x_{1},x_2,\ldots,x_{n}\right):x_{1}\geq x_2\geq\ldots\geq x_{n}> 0\right\}$.
\item[(ii)] $\mathcal{E}_{+}=\left\{\left(x_{1},x_2,\ldots,x_{n}\right):0< x_{1}\leq x_2\leq\ldots\leq x_{n}\right\}$.
\end{enumerate}
\end{n1}
Next, two lemmas are given which will be used to prove our main results. The first one can be obtained by combining  Proposition H2 of Marshall \emph{et al.} (\cite{Maol}, p. 132) and Lemma 3.2 of Kundu \emph{et al.} (\cite{kun1}) while the second one is due to Lemma 3.4 of Kundu \emph{et al.} (\cite{kun1}).
 \begin{l1}\label{l3}
 Let $\varphi({\bf x})=\sum_{i=1}^ng_i(x_i)$ with ${\bf x}\in \mathcal{D}_+$, where $g_i:\mathbb{R}\to\mathbb{R}$ is differentiable, for all $i=1,2,\ldots, n$. 
Then $\varphi(\mf{x})$ is Schur-convex (Schur-concave) on $\mathcal{D}_+$ if, and only if, 
$$g_{i}'(a)\geq\; (resp. \leq)\ g_{i+1}'(b)\;\text{whenever}\;a\geq b,\;\text{for all}\;i=1,2,\ldots,n-1,$$
where $g'(a)=\frac{d g(x)}{dx}\big|_{x=a}$. 
 \end {l1}
\begin{l1}\label{l4}
 Let $\varphi({\bf x})=\sum_{i=1}^ng_i(x_i)$ with ${\bf x}\in \mathcal{E}_+$, where $g_i:\mathbb{R}\to\mathbb{R}$ is differentiable, for all $i=1,2,\ldots, n$. 
Then $\varphi(\mf{x})$ is Schur-convex (Schur-concave) on $\mathcal{E}_+$ if, and only if, 
$$g_{i+1}'(a)\geq\;(resp. \leq)\  g_{i}'(b)\;\text{whenever}\;a\geq b,\;\text{for all}\;i=1,2,\ldots,n-1,$$
where $g'(a)=\frac{d g(x)}{dx}\big|_{x=a}$. 
 \end {l1}

\section{Main Results}
\setcounter{equation}{0}
\hspace{0.3in} For $i=1,2,\ldots,n$, let $X_i$ (resp. $Y_i$) be $n$ independent nonnegative random variables following LL distribution as given in (\ref{e0}).\\
\hspace*{0.3 in} If $F_{n:n}\left(\cdot\right)$ and $G_{n:n}\left(\cdot\right)$ be the distribution functions of $X_{n:n}$ and $Y_{n:n}$ respectively, where $\mbox{\boldmath$\sigma$}=\left(\sigma_1,\sigma_2,\ldots,\sigma_n\right)$, $\mbox{\boldmath$\theta$}=\left(\theta_1,\theta_2, \ldots,\theta_n\right)$, $\mbox{\boldmath$\lambda$}=\left(\lambda_1,\lambda_2,\ldots,\lambda_n\right)$ and $\mbox{\boldmath$\delta$}=\left(\delta_1,\delta_2,\ldots,\delta_n\right)$, then
\begin{equation*}
F_{n:n}\left(x\right)=\prod_{i=1}^n \frac{x^{\sigma_{i}}\left(1+\sigma_{i}\left(\lambda_{i}-\log x\right)\right)}{1+\lambda_{i}\sigma_{i}},
\end{equation*}
and
\begin{equation*}
G_{n:n}\left(x\right)=\prod_{i=1}^n \frac{x^{\theta_{i}}\left(1+\theta_{i}\left(\delta_{i}-\log x\right)\right)}{1+\delta_{i}\theta_{i}}.
\end{equation*}
Again, if $\tilde{r}_{n:n}^{X}$ and $\tilde{r}_{n:n}^{Y}$ are the reversed hazard rate functions of $X_{n:n}$ and $Y_{n:n}$ respectively, then
\begin{equation}
\tilde{r}_{n:n}^X\left(x\right)=\sum_{i=1}^n\frac{\sigma_i}{x}\left(1-\frac{1}{1+\sigma_i\left(\lambda_i-\log x\right)}\right)\label{e1},
\end{equation}
and 
\begin{equation}
\tilde{r}_{n:n}^Y\left(x\right)=\sum_{i=1}^n\frac{\theta_i}{x}\left(1-\frac{1}{1+\theta_i\left(\delta_i-\log x\right)}\right)\label{e2}.
\end{equation}
\hspace*{0.3 in}The following two theorems show that under certain conditions on parameters, there exists reversed hazard rate ordering between $X_{n:n}$ and $Y_{n:n}$.

\begin{t1}\label{th1}
For $i=1,2,\ldots, n$, let $X_i$ and $Y_i$ be two sets of mutually independent random variables with $X_i\sim LL\left(\sigma_i,\lambda_i\right)$ and $Y_i\sim LL\left(\theta_i,\lambda_i\right)$. Further, suppose that
 $\mbox{\boldmath $\sigma$}, \mbox{\boldmath $\theta$}, \mbox{\boldmath $\lambda$}\in \mathcal{D}_+$ or $\mbox{\boldmath $\sigma$}, \mbox{\boldmath $\theta$}, \mbox{\boldmath $\lambda$}\in \mathcal{E}_+$.
 Then, $$\mbox{\boldmath $\sigma$}\stackrel{m}{\succeq}\mbox{\boldmath $\theta$}\;\text{implies}\; X_{n:n}\ge_{rhr}Y_{n:n}.$$
\end{t1}
{\bf Proof:} 
Let $g_{i}(y)=\frac{y}{x}\left(1-\frac{1}{1+y\left(\lambda_i-\log x\right)}\right).$ Differentiating $g_{i}(y)$ with respect to $y$, we get $$g_{i}^{'}(y)=\frac{1}{x}\left(1-\frac{1}{\left(1+y\left(\lambda_i-\log x\right)\right)^2}\right),$$ giving $$g_{i}^{'}(\sigma_i)-g_{i+1}^{'}(\sigma_{i+1})=\frac{\left(1+\sigma_i\left(\lambda_i-\log x\right)\right)^2-\left(1+\sigma_{i+1}\left(\lambda_{i+1}-\log x\right)\right)^2}{x\left(\left(1+\sigma_i\left(\lambda_i-\log x\right)\right)\left(1+\sigma_{i+1}\left(\lambda_{i+1}-\log x\right)\right)\right)^2}.$$ 
So, if $\mbox{\boldmath $\sigma$}, \mbox{\boldmath $\lambda$}\in \mathcal{D}_+ \left(resp.\ \mathcal{E}_+\right)$, then $g_{i}^{'}(\sigma_i)-g_{i+1}^{'}(\sigma_{i+1})\geq \left(\leq\right)0.$ Then, by Lemma \ref{l3} (Lemma \ref{l4}), $\tilde{r}^X_{n:n}\left(x\right)$ is Schur convex in  $\mbox{\boldmath $\sigma$}$, proving the result. $\hfill\Box$\\
The counterexample given below shows that the ascending (descending) order of the components of the scale and shape parameters are necessary for the result of Theorem \ref{th1} to hold. 
\begin{e1}\label{ce2}
Let $X_i\sim LL\left(\sigma_i, \lambda_i\right)$ and $Y_i\sim LL\left(\theta_i, \lambda_i\right), i=1,2,3.$ Now, if $\left(\sigma_1,\sigma_2, \sigma_3\right)=\left(1, 1,5\right)\in \mathcal{E}_+$, $\left(\theta_1,\theta_2, \theta_3\right)=\left(1,2,4\right)\in \mathcal{E}_+$ and $\left(\lambda_1, \lambda_2, \lambda_3\right)=\left(4, 3, 0.2\right)\in \mathcal{D}_+$ are taken, then from Figure \ref{fig1}, it is clear that $\frac{F_{3:3}(x)}{G_{3:3}(x)}$ is not monotone, giving that $X_{3:3}\ngeq_{rhr}Y_{3:3}$, although $\mbox{\boldmath $\sigma$}\stackrel{m}{\succeq}\mbox{\boldmath $\theta$}$. 
\begin{figure}[t]\centering
\includegraphics[height=7 cm]{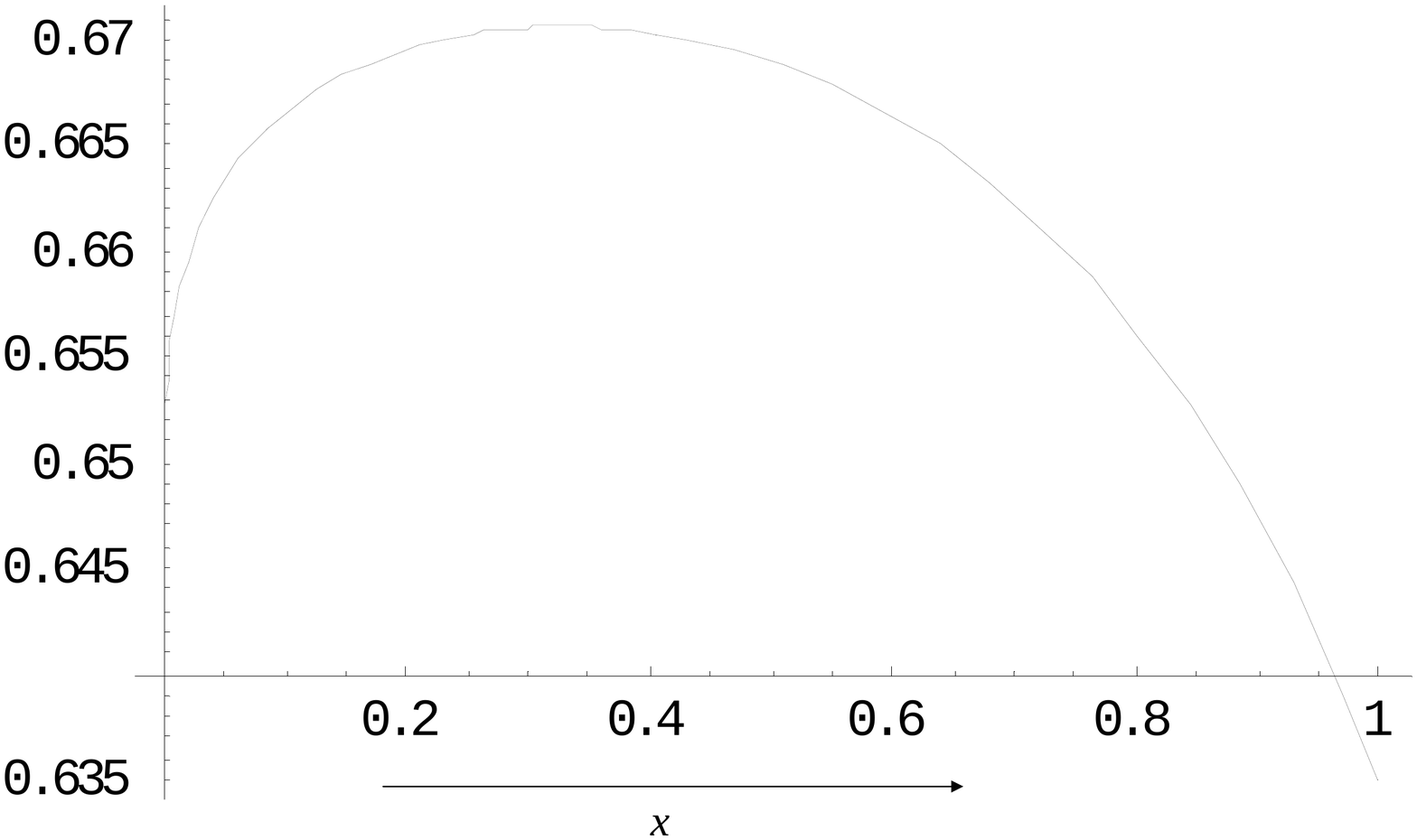}
\caption{\label{fig1} Graph of $\frac{F_{3:3}(x)}{G_{3:3}(x)}$} 
\end{figure}
\end{e1}
\hspace*{0.3in} Theorem \ref{th1} guarantees that for parallel systems of components having independent LL distributed lifetimes with common scale parameter vector, the majorized shape parameter vector leads to larger system's life in the sense of the reversed hazard rate ordering. Now the question arises$-$what will happen if the scale parameter $\mbox{\boldmath $\lambda$}$ majorizes $\mbox{\boldmath $\delta$}$ when the shape parameter vector remains constant? The theorem given below answers that if the order of the components of shape and scale parameter vectors are reversed, then $X_{n:n}$ will be smaller than  $Y_{n:n}$ in reversed hazard rate ordering. 
\begin{t1}\label{th2}
 For $i=1,2,\ldots, n$, let $X_i$ and $Y_i$ be two sets of mutually independent random variables with $X_i\sim LL\left(\sigma_i,\lambda_i\right)$ and $Y_i\sim LL\left(\sigma_i,\delta_i\right)$. Further, suppose that
 $\mbox{\boldmath $\sigma$}\in \mathcal{E}_+$, $\mbox{\boldmath $\lambda$}, \mbox{\boldmath $\delta$}\in \mathcal{D}_+$ or $\mbox{\boldmath $\sigma$}\in \mathcal{D}_+$, $\mbox{\boldmath $\lambda$}, \mbox{\boldmath $\delta$}\in \mathcal{E}_+$.
 Then, $$\mbox{\boldmath $\lambda$}\stackrel{m}{\succeq}\mbox{\boldmath $\delta$}\;\text{implies}\; X_{n:n}\leq_{rhr}Y_{n:n}.$$
\end{t1}
{\bf Proof:} For $i= 1, 2\ldots, n$, let us consider $g_{i}(y)=\frac{\sigma_i}{x}\left(1-\frac{1}{1+\sigma_i\left(y-\log x\right)}\right).$ Differentiating $g_{i}(y)$ with respect to $y$, we get $$g_{i}^{'}(y)=\frac{\sigma_i^2}{x\left(1+\sigma_i\left(y-\log x\right)\right)^2},$$ giving 
\begin{equation*}
\begin{split}
g_{i}^{'}(\lambda_i)-g_{i+1}^{'}(\lambda_{i+1})&\stackrel{sign}{=}\left(\sigma_i^2-\sigma_{i+1}^2\right)+\sigma_i^2\sigma_{i+1}^2\left[\left(\lambda_{i+1}-\log x\right)^2-\left(\lambda_{i}-\log x\right)^2\right]\\&\quad+2\sigma_i\sigma_{i+1}\left[\left(\sigma_i\lambda_{i+1}-\sigma_{i+1}\lambda_{i}\right)-\log x\left(\sigma_i-\sigma_{i+1}\right)\right].
\end{split}
\end{equation*}
So, if $\mbox{\boldmath $\lambda$}\in \mathcal{D}_+\left(resp.\ \mathcal{E}_+\right)$ and $\mbox{\boldmath $\sigma$}\in \mathcal{E}_+\left(resp.\ \mathcal{D}_+\right)$, then $g_{i}^{'}(\lambda_i)-g_{i+1}^{'}(\lambda_{i+1})\leq \left(\geq\right)0.$ So, by Lemma \ref{l3} (Lemma \ref{l4}), $\tilde{r}^X_{n:n}\left(x\right)$ is Schur-concave in  $\mbox{\boldmath $\lambda$}$, proving the result.$\hfill\Box$\\
Next, one counterexample is provided to show that, nothing can be said about reversed hazard rate ordering between $X_{n:n}$ and $Y_{n:n}$ if $\mbox{\boldmath $\lambda$}$ majorizes $\mbox{\boldmath $\delta$}$ and all of $\mbox{\boldmath $\lambda$}$, $\mbox{\boldmath $\delta$}$ and $\mbox{\boldmath $\sigma$}$ are either in $\mathcal{E}_+$ or in $\mathcal{D}_+$.
\begin{e1}\label{e3} 
Let $X_i\sim LL\left(\sigma_i, \lambda_i\right)$ and $Y_i\sim\left(\sigma_i, \delta_i\right), i=1,2,3$. Let $\left(\lambda_1,\lambda_2, \lambda_3\right)=\left(0.1, 0.3, 4.1\right)\in \mathcal{E}_+$ and $\left(\delta_1, \delta_2, \delta_3\right)=\left(0.2, 0.3,4\right)\in \mathcal{E}_+$, giving $\mbox{\boldmath $\lambda$}\stackrel{m}{\succeq}\mbox{\boldmath $\delta$}$. Now, if $\left(\sigma_1,\sigma_2, \sigma_3\right)=\left(0.1, 3,5\right)\in \mathcal{E}_+$ is taken, then Figure \ref{fig2} (a) shows that $\frac{F_{3:3}(x)}{G_{3:3}(x)}$ is increasing in $x$. Again if $\left(\sigma_1,\sigma_2, \sigma_3\right)=\left(2, 3,5\right)\in \mathcal{E}_+$ is taken, then Figure \ref{fig2} (b) shows that $\frac{F_{3:3}(x)}{G_{3:3}(x)}$ is decreasing in $x$. So, it can be concluded that, for all $\mbox{\boldmath $\sigma$}, \mbox{\boldmath$\lambda$}, \mbox{\boldmath$\delta$}\in \mathcal{D}_+ (resp.\ \mathcal{E}_+)$, $\mbox{\boldmath $\lambda$}\stackrel{m}{\succeq}\mbox{\boldmath $\delta$}$ does not always imply $X_{3:3}\leq_{rhr}Y_{3:3}$. 
\begin{figure}[ht]
\centering
\begin{minipage}[b]{0.48\linewidth}
\includegraphics[height=6.5 cm]{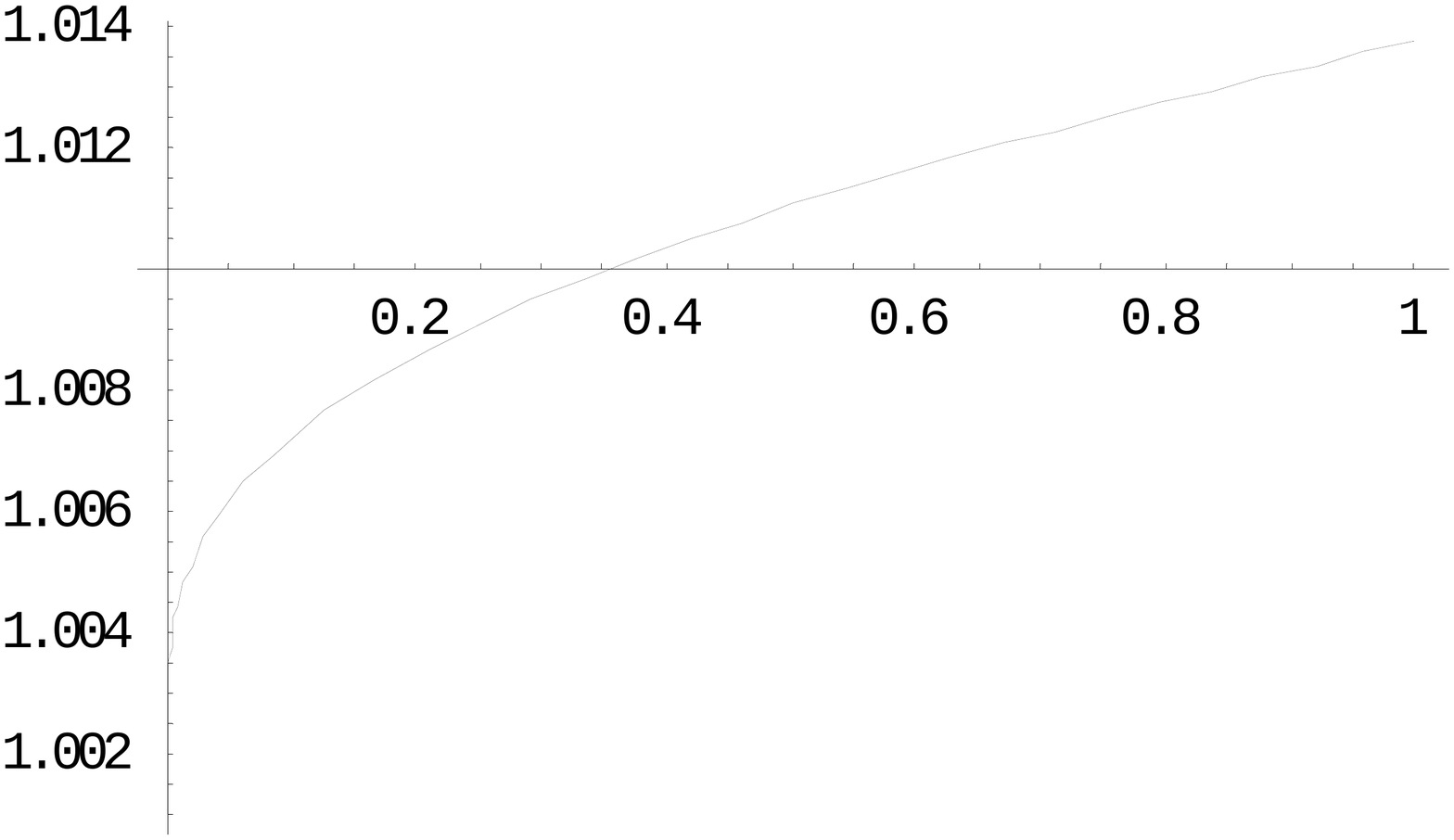}
\centering{(a) For $\left(\sigma_1,\sigma_2, \sigma_3\right)=\left(0.1, 3,5\right)$}
\end{minipage}
\quad
\begin{minipage}[b]{0.48\linewidth}
\includegraphics[height=6.5 cm]{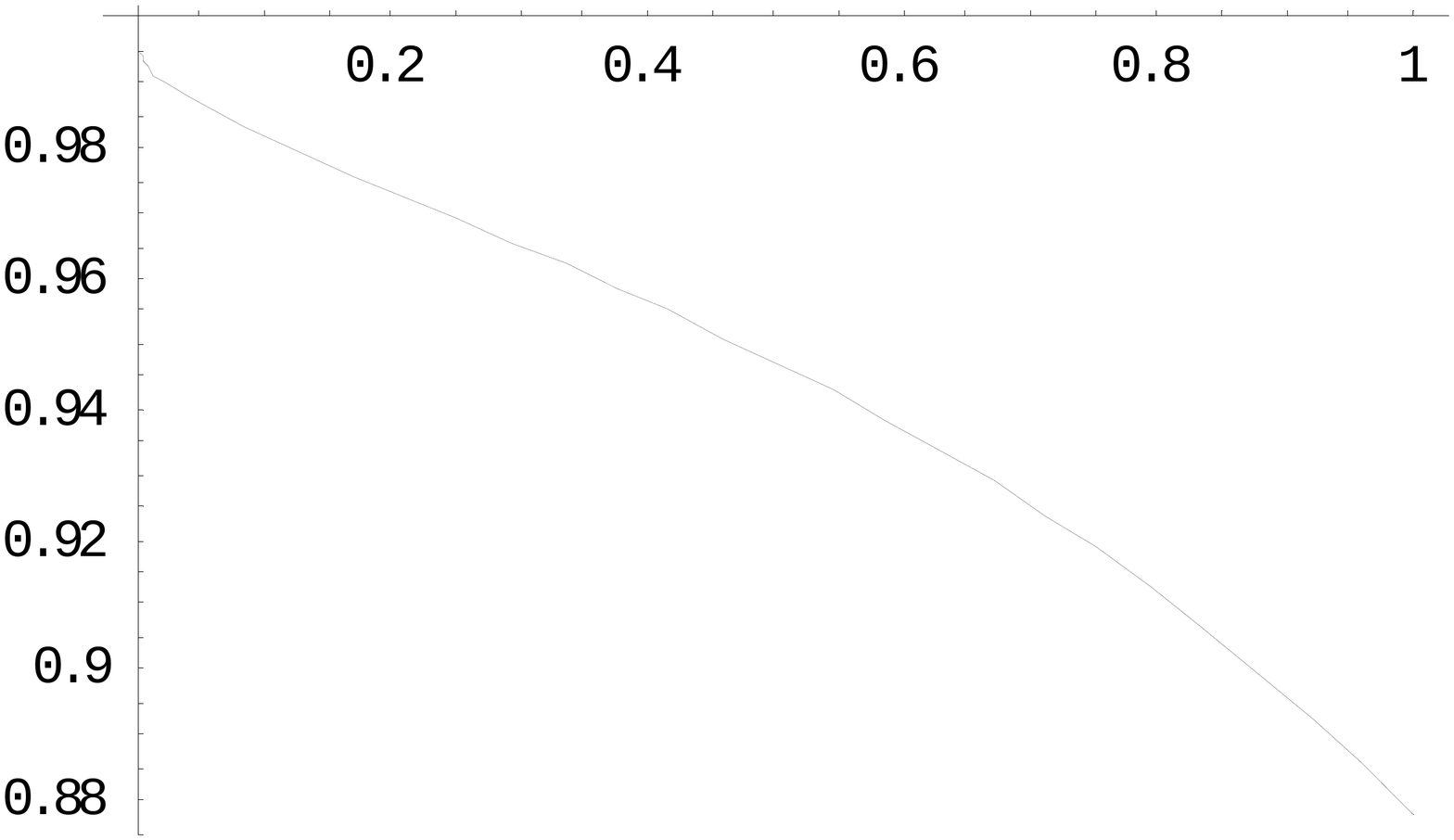}
\centering{(b) For $\left(\sigma_1,\sigma_2, \sigma_3\right)=\left(2, 3,5\right)$}
\end{minipage}
\caption{\label{fig2}Graph of $\frac{F_{3:3}(x)}{G_{3:3}(x)}$}
\end{figure}
\end{e1} 
\hspace*{0.3in} The following theorem shows that depending upon certain conditions, majorization order of the shape parameters implies likelihood ratio ordering between $X_{n:n}$ and $Y_{n:n}$.
\begin{t1}\label{th3}
For $i=1,2,\ldots, n$, let $X_i$ and $Y_i$ be two sets of mutually independent random variables with $X_i\sim LL\left(\sigma_i,\lambda_i\right)$ and $Y_i\sim LL\left(\theta_i,\lambda_i\right)$. Further, suppose that
 $\mbox{\boldmath $\sigma$}, \mbox{\boldmath $\theta$}, \mbox{\boldmath $\lambda$}\in \mathcal{D}_+$ or $\mbox{\boldmath $\sigma$}, \mbox{\boldmath $\theta$}, \mbox{\boldmath $\lambda$}\in \mathcal{E}_+$.
 Then, if $\lambda_i\sigma_i>1/2$, $$\mbox{\boldmath $\sigma$}\stackrel{m}{\succeq}\mbox{\boldmath $\theta$}\;\text{implies}\; X_{n:n}\ge_{lr}Y_{n:n}.$$ 
\end{t1}
{\bf Proof:} %Let, $f_{n:n}(x)$ and  $g_{n:n}(x)$ be the density function of $X_{n:n}$ and $Y_{n:n}$ respectively. Then, from (\ref{e1}) and (\ref{e2}) it can be written that, 
In view of theorem \ref{th1} and using (3.1) and (3.2), here we have only to show that 
\begin{eqnarray*}
\frac{\tilde{r}_{n:n}^X\left(x\right)}{\tilde{r}_{n:n}^Y\left(x\right)}&=&\frac{\sum_{k=1}^nu_k\left(\sigma_k,x\right)}{\sum_{k=1}^nu_k\left(\theta_k,x\right)}\\
&=& \eta(x) (say),
\end{eqnarray*} is increasing in $x$, where $u_k(y,x)=\frac{y^{2}\left(\lambda_k-\log x\right)}{1+y\left(\lambda_k-\log x\right)}$. Now, differentiating $\eta(x)$ with respect to $x$,
\begin{eqnarray*}
\eta^{'}(x)&\stackrel{sign}{=}&\sum_{k=1}^n\frac{\partial u_k\left(\sigma_k,x\right)}{\partial x}\sum_{k=1}^nu_k\left(\theta_k,x\right)-\sum_{k=1}^n\frac{\partial u_k\left(\theta_k,x\right)}{\partial x}\sum_{k=1}^nu_k\left(\sigma_k,x\right)\\&=&-h\left(\mbox{\boldmath$\sigma$},x\right)\sum_{k=1}^nu_k\left(\theta_k,x\right)+h\left(\mbox{\boldmath$\theta$},x\right)\sum_{k=1}^nu_k\left(\sigma_k,x\right),
\end{eqnarray*} 
where $$ h(\mbox{\boldmath$\sigma$}, x)=-\sum_{k=1}^n\frac{\partial u_k\left(\sigma_k,x\right)}{\partial x}=\frac{1}{x}\sum_{k=1}^n\frac{\sigma_k^2}{\left(1+\sigma_k\left(\lambda_k-\log x\right)\right)^2}
$$
and
$$ h(\mbox{\boldmath$\theta$}, x)=-\sum_{k=1}^n\frac{\partial u_k\left(\theta_k,x\right)}{\partial x}=\frac{1}{x}\sum_{k=1}^n\frac{\theta_k^2}{\left(1+\theta_k\left(\lambda_k-\log x\right)\right)^2}.
$$ 
Thus, to show that $\eta(x)$ is increasing in $x$, we have only to show that $$\psi\left(\mbox{\boldmath$\sigma$}, x\right)=\frac{h(\mbox{\boldmath$\sigma$}, x)}{\sum_{k=1}^nu_k\left(\sigma_k,x\right)}$$
is  Schur-concave in $\mbox{\boldmath$\sigma$}$. \\
Now, as 
$$\frac{\partial h(\mbox{\boldmath$\sigma$}, x)}{\partial \sigma_i}=\frac{1}{x}.\frac{2\sigma_i}{\left(1+\sigma_i(\lambda_i-\log x)\right)^3}$$
and
$$\frac{\partial }{\partial \sigma_i}\left[\sum_{k=1}^n u_k\left(\sigma_k,x\right)\right]=1-\frac{1}{\left(1+\sigma_i(\lambda_i-\log x)\right)^2},$$
then
\begin{eqnarray*}
\frac{\partial \psi}{\partial\sigma_i}&\stackrel{sign}{=}&\frac{2\sigma_i}{\left(1+\sigma_i\left(\lambda_i-\log x\right)\right)^3}\sum_{k=1}^n u_k\left(\sigma_k,x\right)-x.h(\mbox{\boldmath$\sigma$}, x)\left(1-\frac{1}{\left(1+\sigma_i(\lambda_i-\log x)\right)^2}\right).
\end{eqnarray*}
So, if $\mbox{\boldmath $\sigma$}, \mbox{\boldmath $\lambda$}\in \mathcal{D}_+ \left(resp \ \in \mathcal{E}_+\right)$, i.e., for $i\leq j$ if $\sigma_i\geq\sigma_j$ and $\lambda_i\geq\lambda_j$ $\left(\sigma_i\leq\sigma_j, \lambda_i\leq\lambda_j\right)$, then noticing the fact that $\frac{1}{\left(1+\sigma_i\left(\lambda_i-\log x\right)\right)^2}$ is decreasing in $\sigma_i$ as well as in $\lambda_i,$ it can be written that $$\frac{1}{\left(1+\sigma_i\left(\lambda_i-\log x\right)\right)^2}\leq (\geq) \frac{1}{\left(1+\sigma_j\left(\lambda_i-\log x\right)\right)^2}\leq (\geq) \frac{1}{\left(1+\sigma_j\left(\lambda_j-\log x\right)\right)^2}.$$
Again, as $\sigma_i\lambda_i>\frac{1}{2}$ implying $\sigma_i\left(\lambda_i-\log x\right)>\frac{1}{2}$ for all $0<x<1$, then  
$$\frac{\partial}{\partial\sigma_i}\left(\frac{\sigma_i}{\left(1+\sigma_i\left(\lambda_i-\log x\right)\right)^3}\right)=\frac{1-2\sigma_i\left(\lambda_i-\log x\right)}{\left(1+\sigma_i\left(\lambda_i-\log x\right)\right)^4}<0,$$ proving that $\frac{\sigma_i}{\left(1+\sigma_i\left(\lambda_i-\log x\right)\right)^3}$ is decreasing in $\sigma_i$. Again, it is also decreasing in $\lambda_i$. Thus, for all $\sigma_i\geq\sigma_j$ and $\lambda_i\geq\lambda_j$ $\left(\sigma_i\leq\sigma_j, \lambda_i\leq\lambda_j\right)$,
$$\frac{\sigma_i}{\left(1+\sigma_i\left(\lambda_i-\log x\right)\right)^3}\leq (\geq)\frac{\sigma_j}{\left(1+\sigma_j\left(\lambda_i-\log x\right)\right)^3}\leq (\geq)\frac{\sigma_j}{\left(1+\sigma_j\left(\lambda_j-\log x\right)\right)^3}.$$
So, for all $i\leq j$
\begin{equation*}
\begin{split}
\frac{\partial\psi}{\partial\sigma_i}-\frac{\partial\psi}{\partial\sigma_j}&\stackrel{sign}{=}\sum_{k=1}^n\frac{\sigma_k^2\left(\lambda_k-\log x\right)}{1+\sigma_k\left(\lambda_k-\log x\right)}\left[\frac{2\sigma_i}{\left(1+\sigma_i\left(\lambda_i-\log x\right)\right)^3}-\frac{2\sigma_j}{\left(1+\sigma_j\left(\lambda_j-\log x\right)\right)^3}\right]\\
&\quad+\sum_{k=1}^n\frac{\sigma_k^2}{\left(1+\sigma_k\left(\lambda_k-\log x\right)\right)^2}\left[\frac{1}{\left(1+\sigma_i\left(\lambda_i-\log x\right)\right)^2}-\frac{1}{\left(1+\sigma_j\left(\lambda_j-\log x\right)\right)^2}\right]\\
&\leq (\geq) 0.
\end{split}
\end{equation*}
Thus the result follows from Lemma 3.1 (Lemma 3.3) of Kundu \emph{et al.} (\cite{kun1}).$\hfill\Box$\\
\hspace*{0.3in} Although Theorem \ref{th3} holds under a sufficient condition for two $n$ component systems, the next theorem shows that no such condition is required for these systems having multiple-outlier LL model if the scale parameter vectors of these systems are common.
\begin{t1}\label{th5}
 For $i=1,2,...,n$, let $X_i$ and $Y_i$ be two sets of independent random variables each following the multiple-outlier EW model such that $X_i\sim LL\left(\sigma,\lambda\right)$ and $Y_i\sim LL\left(\theta,\lambda\right)$ for $i=1,2,\ldots,n_1$,
 $X_i\sim LL\left(\sigma^*,\lambda^*\right)$ and $Y_i\sim LL\left(\theta^*,\lambda^*\right)$ for $i=n_1+1,n_1+2,\ldots,n_1+n_2(=n)$ If $$(\underbrace{\sigma,\sigma,\ldots,\sigma,}_{n_1} \underbrace{\sigma^*,\sigma^*,\ldots,\sigma^*}_{n_2})\stackrel{m}{\succeq}
(\underbrace{\theta,\theta,\ldots,\theta,}_{n_1} \underbrace{\theta^*,\theta^*,\ldots,\theta^*}_{n_2})$$ 
and either $\{\sigma\ge\sigma^*, \theta\ge\theta^*, \lambda\ge\lambda^*\}$ or $\{\sigma\leq\sigma^*, \theta\leq\theta^*, \lambda\leq\lambda^*\}$ then $ X_{n:n}\ge_{lr}Y_{n:n}$. 
\end{t1}
{\bf Proof:} Following Theorem \ref{th3} and 
%we get
%\begin{eqnarray*}
%\frac{\tilde{r}_{n:n}^{X}(x)}{\tilde{r}_{n:n}^{Y}(x)}&=&\frac{\sum_{i=1}^n\frac{\sigma_i^{2}}{1+\sigma_i\left(\lambda_i-\log x\right)}}{\sum_{i=1}^n\frac{\theta_i^{2}}{1+\theta_i\left(\lambda-\log x\right)}}\\&=& \frac{\sum_{i=1}^{n} \phi(\sigma_i, x)}{\sum_{i=1}^{n} \phi(\theta_i, x)}x)\\&=& \eta_{1}(x) (say),
%\end{eqnarray*}
in view of Theorem \ref{th1}, we have only to show that 
%$\eta_{1}(x)$ is increasing in $x.$ Now,  
%\begin{equation*}
%\begin{split}
 %\eta'(x)&\stackrel{\rm sign}=\left(\phi^{'}(\sigma_i, x)F_{n:n}(x)+ \phi(\sigma_i, x)f_{n:n}(x)\right)\left(\phi(\theta_i, x)G_{n:n}(x)\right)
 %\\&\quad-\left(\phi^{'}(\theta_i, x)G_{n:n}(x)+ \phi(\theta_i, x)g_{n:n}(x)\right)\left(\phi(\sigma_i, x)F_{n:n}(x)\right).
%\end{split}
%\end{equation*}
%$\eta_{1}(x)$ is increasing in $x\ge 0$ if 
$$\psi_{1}(\mbox{\boldmath$\sigma$},x)=\frac{\sum_{k=1}^n\frac{\sigma_k^{2}}{\left(1+\sigma_k(\lambda_k-\log x)\right)^{2}}}{\sum_{k=1}^n\frac{\sigma_k^{2}(\lambda_k-\log x)}{1+\sigma_k(\lambda_k-\log x)}}$$
is Schur-concave in $\mbox{\boldmath$\sigma$}$.\\
% or equivalently 
%$$\Psi_{1}(\mbox{\boldmath$\sigma$})=\frac{\sum_{i=1}^n\frac{\xi_i^{3}}{\left(1+\xi_i\right)^{2}}}{\sum_{i=1}^n\frac{\xi_i^{2}}{1+\xi_i}}$$ is Schur-convex in $(\mbox{\boldmath$\xi$}),$ where $\xi_i=\sigma_i(\lambda-\log x)(>0)$ as earlier with $\xi_i=\xi_1=\sigma(\lambda-\log x)$ for $i=1, 2, \ldots n_1$ and $\xi_i=\xi_2=\sigma^{*}(\lambda-\log x)$ for $i=n_1+1, n_1+2, \ldots n_1+n_2=n$.\\
\hspace*{0.3 in} Now, three cases may arise:\\
$Case (i)$ If $1\leq i<j\leq n_1$, $i.e.$, if $\sigma_i=\sigma_j=\sigma$ and $\lambda_i=\lambda_j=\lambda$, then $\frac{\partial \psi_{1}}{\partial \sigma_i}-\frac{\partial \psi_{1}}{\partial \sigma_j}=0.$ \\ 
$Case (ii)$ If $n_1+1\leq i<j\leq n$, $i.e.$, if $\sigma_i=\sigma_j=\sigma^*$ and $\lambda_i=\lambda_j=\lambda^*$, then $\frac{\partial \Psi}{\partial \sigma_i}-\frac{\partial \Psi}{\partial \sigma_j}=0.$ \\ 
$Case (iii)$ If $1\leq i\leq n_1$ and $n_1+1\leq j\leq n$, then $\sigma_i=\sigma$, $\lambda_i=\lambda$ and $\sigma_j=\sigma^*$, $\lambda_i=\lambda^*$. It can be easily shown that 
\begin{equation*}
\begin{split}
\frac{\partial \psi_1}{\partial \sigma_i}-\frac{\partial \psi_1}{\partial \sigma_j}&\stackrel{sign}{=}\left(\frac{n_1\sigma^{2}}{(1+\xi_1)^{2}}+\frac{n_2\sigma^{*2}}{(1+\xi_2)^{2}}\right)\left(\frac{\xi_2^{2}}{(1+\xi_2)^{2}}-\frac{\xi_1^{2}}{(1+\xi_1)^{2}}\right)\\&\quad+\left(\frac{\sigma\xi_2}{(1+\xi_1)}-\frac{\sigma^*\xi_1}{(1+\xi_2)}\right)\left(\frac{2n_1\sigma}{(1+\xi_2)^2(1+\xi_1)}+\frac{2n_2\sigma^*}{(1+\xi_1)^2(1+\xi_2)}\right).
\end{split}
\end{equation*}
where $\xi_1=\sigma(\lambda-\log x)$ and $\xi_2=\sigma^{*}(\lambda^*-\log x)$. Now, as $\sigma \geq (\leq) \sigma^*$ and $\lambda \geq (\leq) \lambda^*$, implying that $\sigma(\lambda-\log x) \geq (\leq) \sigma^*(\lambda^*-\log x)$ $i.e.$ $\xi_1\geq (\leq)\xi_2$, and moreover, $\frac{\xi}{1+\xi}=1-\frac{1}{1+\xi}$ is increasing in $\xi$, then $\frac{\xi_2^{2}}{(1+\xi_2)^{2}}\leq (\geq)\frac{\xi_1^{2}}{(1+\xi_1)^{2}}$. Again,
\begin{eqnarray*}
\frac{\sigma\xi_2}{1+\xi_1}-\frac{\sigma^*\xi_1}{1+\xi_2}&=&\frac{\sigma\sigma^*\left\{(\lambda^*-\log x)(1+\sigma^*(\lambda^*-\log x))-(\lambda-\log x)(1+\sigma(\lambda-\log x))\right\}}{\left(1+\xi_1\right)\left(1+\xi_2\right)}\\
&\le(\ge)& 0.
\end{eqnarray*} 
So, by Lemma 3.1 (Lemma 3.3) of Kundu \emph{et al.} (\cite{kun1}), the result is proved. $\hfill\Box$\\ 
\hspace*{0.3in} Theorem \ref{th3} guarantees that, for two $n$ component  parallel systems (with a sufficient condition) having independent LL distributed lifetimes with a common scale parameter vector, the majorized shape parameter vector leads to greater system's lifetime in the sense of likelihood ratio order. The next theorem states that the majorized scale parameter vector leads to smaller system's lifetime in the sense of likelihood ratio order when the shape parameter vector of these two $n$-component parallel systems are common. 
\begin{t1}\label{th6}
For $i=1,2,\ldots, n$, let $X_i$ and $Y_i$ be two sets of mutually independent random variables with $X_i\sim LL\left(\sigma_i,\lambda_i\right)$ and $Y_i\sim LL\left(\sigma_i,\delta_i\right)$. Further, suppose that
 $\mbox{\boldmath $\sigma$}\in \mathcal{E}_+$, $\mbox{\boldmath $\lambda$}, \mbox{\boldmath $\delta$}\in \mathcal{D}_+$ or $\mbox{\boldmath $\sigma$}\in \mathcal{D}_+$, $\mbox{\boldmath $\lambda$}, \mbox{\boldmath $\delta$}\in \mathcal{E}_+$.
 Then, $$\mbox{\boldmath $\lambda$}\stackrel{m}{\succeq}\mbox{\boldmath $\delta$}\;\text{implies}\; X_{n:n}\leq_{lr}Y_{n:n}.$$ 
\end{t1}
{\bf Proof:} In view of Theorem \ref{th2} and using (\ref{e1}) and (\ref{e2}), we are to prove that
\begin{equation*}
\eta_{1}(x)=\frac{\sum_{k=1}^n\frac{\sigma_k}{x}\left(1-\frac{1}{1+\sigma_k(\lambda_k-\log x)}\right)}{\sum_{k=1}^n\frac{\sigma_k}{x}\left(1-\frac{1}{1+\sigma_k(\delta_k-\log x)}\right)}
\end{equation*}
is decreasing in $x$ $i.e.$ to prove that  
$$\psi_{2}(\mbox{\boldmath$\lambda$},x)=\frac{\sum_{k=1}^n\frac{\sigma_k^2}{\left(1+\sigma_k\left(\lambda_k-\log x\right)\right)^2}}{\sum_{k=1}^n\frac{\sigma_k^2\left(\lambda_k-\log x\right)}{\left(1+\sigma_k\left(\lambda_k-\log x\right)\right)}}$$
is Schur-convex in $\mbox{\boldmath$\lambda$}$. Now, 
$$\frac{\partial\psi_2}{\partial\lambda_i}\stackrel{sign}{=}-\frac{2\sigma_i^3}{\left(1+\sigma_i\left(\lambda_i-\log x\right)\right)^3}\sum_{k=1}^n\frac{\sigma_k^2\left(\lambda_k-\log x\right)}{\left(1+\sigma_k\left(\lambda_k-\log x\right)\right)}-\frac{\sigma_i^2}{\left(1+\sigma_i\left(\lambda_i-\log x\right)\right)^2}\sum_{k=1}^n\frac{\sigma_k^2}{\left(1+\sigma_k\left(\lambda_k-\log x\right)\right)^2}.$$  
So, by noticing the fact that
$$\frac{\partial}{\partial\sigma_i}\left[\frac{\sigma_i}{\left(1+\sigma_i\left(\lambda_i-\log x\right)\right)}\right]=\frac{1}{\left(1+\sigma_i\left(\lambda_i-\log x\right)\right)^2}>0,$$
giving that $\frac{\sigma_i}{\left(1+\sigma_i\left(\lambda_i-\log x\right)\right)}$ is increasing in $\sigma_i$, $\mbox{\boldmath $\lambda$}\in \mathcal{D}_+ \left(resp. \mathcal{E}_+\right)$ and $\mbox{\boldmath $\sigma$}\in \mathcal{E}_+ \left(resp. \mathcal{D}_+\right)$, $i.e.$ for all $i\le j$ $\lambda_i\geq(\leq)\lambda_j$ and $\sigma_i\leq(\geq)\sigma_j$ gives
$$\frac{\sigma_i^3}{\left(1+\sigma_i\left(\lambda_i-\log x\right)\right)^3}\leq(\geq)\frac{\sigma_j^3}{\left(1+\sigma_j\left(\lambda_i-\log x\right)\right)^3}\leq(\geq)\frac{\sigma_j^3}{\left(1+\sigma_j\left(\lambda_j-\log x\right)\right)^3}$$
and
$$\frac{\sigma_i^2}{\left(1+\sigma_i\left(\lambda_i-\log x\right)\right)^2}\leq(\geq)\frac{\sigma_j^2}{\left(1+\sigma_j\left(\lambda_i-\log x\right)\right)^2}\leq(\geq)\frac{\sigma_j^2}{\left(1+\sigma_j\left(\lambda_j-\log x\right)\right)^2}.$$
So,
$$\frac{\partial\psi_2}{\partial\lambda_i}-\frac{\partial\psi_2}{\partial\lambda_j}\geq(\leq)0.$$
Thus the result follows from Lemma 3.1 (Lemma 3.3) of Kundu \emph{et al.} (\cite{kun1}).$\hfill\Box$\\

\end{document}